# Calculation and verification of neutron irradiation damage with differential cross sections

Shengli Chen[1,3,*], David Bernard[1], Pierre Tamagno[1], Jean Tommasi[1], Stéphane Bourganel[2], Gilles Noguère[1], Cyrille De Saint Jean[1]

[1] CEA, Cadarache, DEN/DER/SPRC/LEPh, 13108 Saint Paul Les Durance, France
[2] CEA, Saclay, DEN/DM2S/SERMA/LPEC, 91191 Gif sur Yvette, France
[3] Université Grenoble Alpes, I-MEP2, 38402 Saint Martin d'Hères, France
[*] Corresponding author: shengli.chen@cea.fr

**Abstract**

The Displacement per Atom (DPA) rate is conventionally computed with DPA cross sections in reactor applications. The method of direct calculation with energy-angular distributions given in the Center of Mass (CM) frame is proposed and recommended in the present work. The methods for refining and verifying the calculations of DPA cross sections are proposed: (i) Gauss-Legendre-Quadrature-based Piecewise Integration (GLQPI) for ensuring the numeric convergence of integral over emission angle due to the discontinuity of integrand; (ii) verification of the convergence for trapezoidal integration over the secondary energy; (iii) interpolation of double-differential cross sections. For $^{56}$Fe of JEFF-3.1.1, the current numeric integration over emission angle is shown not convergent, whereas the direct trapezoidal over the secondary energy and the direct interpolation of energy-angle-integrated damage are shown accurate. On the other hand, it is shown that the DPA cross sections are overestimated if isotropic angular distributions are assumed. However, the DPA cross section is not sensitive to the high-order Legendre polynomials because the former is an angle-integrated quantity. Numerical results of neutron elastic scattering show that 2 orders of Legendre polynomials can give the DPA rates of $^{56}$Fe within 0.5% overestimation for fission reactors, while 4 orders are required for fusion reactors. For neutron inelastic scatterings-induced DPA, the first order Legendre polynomial is sufficient for both fission and fusion reactors.

**Keywords:** Displacement per Atom, Differential cross section, Gauss-Legendre quadrature, Legendre polynomial, $^{56}$Fe

## 1. Introduction

In nuclear industry, the neutron embrittlement is one of the major material challenges of the Reactor Pressure Vessel (RPV) [1]. When an atom in a material is knocked-on by a kinematic particle, a vacancy and a corresponding interstitial are formed in the lattice. The Primary Knock-on Atom (PKA) can induce a displacement cascade, which leads to more crystallographic defects in the material. The Displacement per Atom (DPA) intends to estimate the average displacements of each atom under irradiation. It is one of the key parameters to evaluate the irradiated damage. Many models have been developed to calculate DPA using the energy of PKA as a major parameter, such as Molecular Dynamics (MD) simulation [2], Binary Collision Approximation (BCA) [3], Density Functional Theory (DFT) [4], and some universal formulae summarized in Section 2.1.

The typical method of DPA calculation applied in nuclear reactors is the generation of DPA cross sections through the processing code NJOY [5]. The DPA rates can be calculated with the DPA cross sections and the spectra of incident particles computed through the transport codes [6]. According to the nuclear data given in Evaluated Nuclear Data Files (ENDF), the recoil energy of PKA is conventionally calculated through the angular distribution (i.e. differential cross section) of the emission particle for discrete reactions and energy-angular distribution (i.e. double-differential cross section) for continuum reactions. The DPA cross sections are computed with the integral of the damage energy versus the emission angle (and secondary energy for continuum reactions). The present work proposes the methods for investigating the accuracy of numeric integrals used in the calculations of DPA cross sections.

The angular distributions for discrete reactions are conventionally given in the Center-of-Mass (CM) frame, while the energy-angular distributions for continuum reactions are conventionally given in the Laboratory (Lab) frame. The corresponding PKA energies are well developed for these cases (summarized in Sections 2.2 and 2.3) and used in NJOY [5]. However, the energy-angular distributions are often given in the CM frame in ENDF, such as $^{56}$Fe in JEFF-3.1.1 [7]. Three methods for calculating damage cross sections with energy-angular distributions given in the CM frame are presented and compared in Section 2.4. Sections 3.1 and 3.2 show the methods and numeric results for refining numeric integrations over emission angle and secondary energy, respectively. On the other hand, because the energy-angular distributions are tabulated on coarse meshes of incident energy given in ENDF, Section 3.3 shows different methods of DPA cross sections calculations between two given neighbor incident energies, including the direct interpolation of energy-angle-integrated damage energy, the standard and the present work proposed improved methods for interpolation of double-differential cross sections.

The high-order Legendre polynomials are commonly used to describe the anisotropy of angular distributions. The influence of the anisotropic angular distribution on PKA energy was shown in Refs. [8], [9]. On the other hand, Jouanne showed that the first order Legendre polynomial of the angular distribution of $^{56}$Fe is almost sufficient to determine the neutron fluence on the iron bulk from 5 cm to 1.2 m [10]. Therefore, we investigate the importance of high-order Legendre polynomials of angular distribution for DPA calculations in Section 4. The examples on $^{56}$Fe are taken to show the numerical results because of its high abundance in iron-based steels, which are used in RPV in Light Water Reactor (LWR), fuel cladding in Fast Reactor (FR), and candidate fuel cladding in Accident Tolerant Fuel (ATF) [11], [12]. The numerical applications in this paper are mainly based on the nuclear data of JEFF-3.1.1 [7], which is widely used and qualified by CEA, EDF, and FRAMATOME.

## 2. From nuclear data to DPA cross sections

2.1. Summary of different DPA formulae

Many models and empirical formulae have been developed to calculate the DPA in materials. Kinchin and Pease proposed a formula (KP) to calculate the number of displaced atoms induced by a PKA in 1955 [13]. In this model, the PKA cannot produce atomic displacement if the PKA energy $E_{PKA} < E_d$, where $E_d$ is the averaged threshold energy of atomic displacement. Different definitions of the threshold energy and corresponding values for Fe can be found in the Nordlund's work [14]. The commonly used average threshold energy and the value proposed by the ASTM for the iron is 40 eV [15]. When $E_d < E_{PKA} < 2E_d$, one atom is displaced. Once the PKA energy is higher than the ionization energy of the target atom ($E_I$), the excess PKA energy is supposed to be transferred to electrons. The equivalent kinetic energy of PKA is thus equal to $E_I$. The KP formula is mathematically expressed as:

$$N(E_{PKA}) = \begin{cases} 0, & 0 < E_{PKA} < E_d \\ 1, & E_d < E_{PKA} < 2E_d \\ \frac{E_{PKA}}{2E_d}, & 2E_d < E_{PKA} < E_I \\ \frac{E_I}{2E_d}, & E_I < E_{PKA} \end{cases} \quad (1)$$

Using the form of KP formula, Norgett, Robinson, and Torrens proposed the NRT model in 1975 [16]. The NRT formula uses Lindhard's damage energy [17] with Robinson's analytic fitting [18]:

$$N(E_a) = \begin{cases} 0, & 0 < E_a < E_d \\ 1, & E_d < E_a < 2E_d/0.8 \\ \frac{0.8E_a}{2E_d}, & 2E_d/0.8 < E_a \end{cases} \quad (2)$$

where $E_a$ is the energy available to create displacement of atoms by collision, called as damage energy, 0.8 is the displacement efficiency obtained by the BCA by Robinson and Torrens [3]. $E_a = E_{PKA} \times P(E_{PKA}/E_L)$, where $P$ is the partition function which describes the fraction of $E_{PKA}$ left in atomic motion [17], [18]:

$$P(\varepsilon) = 1/[1 + k(3.4008\varepsilon^{1/6} + 0.40244\varepsilon^{3/4} + \varepsilon)] \quad (3)$$

where $\varepsilon = E_{PKA}/E_L$ with $E_L = 86.931Z^{7/3}$ eV, $k = 0.133745Z^{2/3}A^{-1/2}$, $Z$ and $A$ are atomic number and atomic mass number, respectively. It is noticeable that the Lindhard's equation is proposed for PKA energy below $24.9AZ^{4/3}$ keV [17], so the NRT-DPA formula is valid for PKA energy lower than this value.

However, the overestimation of DPA in the NRT model is found in 1977 with experimental data for copper and silver [19]. One of the issues in the NRT model is that the in-cascade recombination of displaced atoms is neglected. Taking this effect into account, the Athermal Recombination-Corrected (ARC)-DPA is proposed by Nordlund et al. [20]. The relative damage efficiency $\xi$ defines the ratio of the "true" number of Frenkel Pairs ($N_{FP}$) to the $N_{FP}$ calculated with NRT formula. Its expression is based on the fact that $N_{FP}$ tends to $a'E_a$ when $E_a$ tends to infinity and $N_{FP}$ tends to $c'E_a^{0.8}$ at low energy [20]. Therefore, the ARC-DPA formula is given by:

$$N(E_a) = \begin{cases} 0, & 0 < E_a < E_d \\ 1, & E_d < E_a < 2E_d/0.8 \\ \frac{0.8E_a}{2E_d}\xi(E_a), & 2E_d/0.8 < E_a \end{cases} \quad (4)$$

where

$$\xi(E_a) = (1-c) \times \left[0.8\frac{E_a}{2E_d}\right]^b + c \quad (5)$$

The coefficients $b$ and $c$ are determined by fitting experimental data or molecular dynamics simulation results. For Fe isotopes, $b = -0.568$ and $c = -0.286$ [21].

2.2. DPA cross sections and angular distribution

Figure 1 shows the schemes of the collision in the Laboratory (Lab) and Center of Mass (CM) frames. The incident and emitted kinetic energies are referred to $E$ and $E'$ in the Lab frame, respectively. $E_R$ stands for the recoil energy of the PKA in the Lab frame. $m$ and $v_1$ ($m'$ and $u_1$) are the mass and velocity of the incident (outgoing) particle in the CM frame, respectively. $M$ and $v_2$ ($M'$ and $u_2$) are the mass and velocity of recoil particle before (after) the collision in the CM frame, respectively.

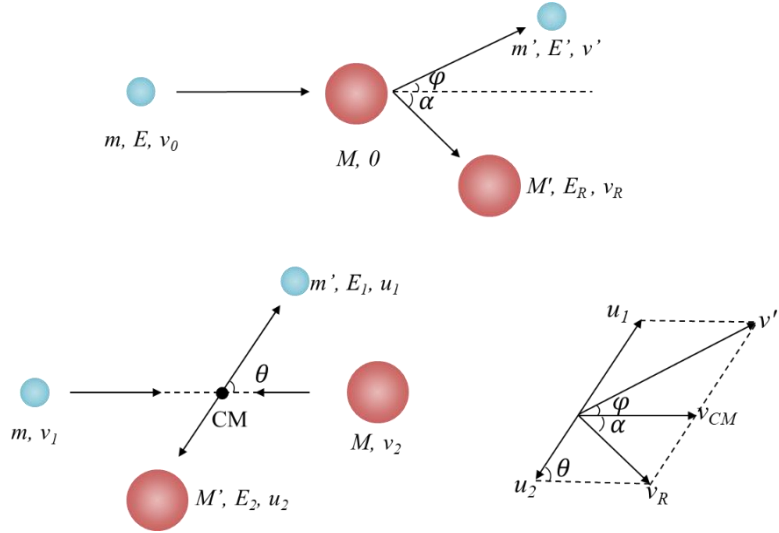

Figure 1. Schemes of the collision in the Laboratory (upper) and Center of Mass (lower) frames

The relativistic effect is negligible for DPA calculations with incident energy lower than 20 MeV [22], [23]. The following studies are based on the classical mechanism. The conservation of energy before and after the collision in the CM frame conducts to:

$$m'c^2 + M'c^2 + \frac{1}{2}m'u_1^2 + \frac{1}{2}M'u_2^2 = mc^2 + Mc^2 + \frac{1}{2}mv_1^2 + \frac{1}{2}Mv_2^2 + Q' \quad (6)$$

where $v_1 = v_0 - v_{CM}$ and $v_2 = v_{CM}$ are used because the thermal vibration of the target has no influence on DPA calculation [6]. $Q'$ is the reaction energy. The total energy change during the collision is $Q = Q' - [(m' + M') - (m + M)]c^2$, which is then transferred into the excitation energy of the recoil nucleus. Transforming the recoil velocity from the CM to the Lab frame:

$$v_R^2 = u_2^2 + v_{CM}^2 - 2u_2 v_{CM} \cos\theta \quad (7)$$

where $\theta$ is the scattering angle of the emitted particle in the CM frame.

The conservation of momentum points out:

$$(m + M)v_{CM} = mv_0 \qquad (8)$$

where $v_0$ and $v_{CM}$ are the initial velocity of the incident particle and the velocity of the Center of Mass in the Lab frame, respectively. The momentum in the CM frame is always null. Hence,

$$m'u_1 = M'u_2 \qquad (9)$$

As a matter of fact, $(m' + M')/(m + M) = 1$ is numerically valid even though quite small percentage of the mass is reduced during the nuclear reactions. Defining the "effective mass" $R(E)$ as:

$$R(E) = \sqrt{1 + \frac{(m+M)Q}{ME}}. \qquad (10)$$

One obtains:

$$E_R = \frac{m'ME}{(m+M)^2}\left[\frac{mM'}{m'M} - 2R(E)\sqrt{\frac{mM'}{m'M}}\mu + R(E)^2\right] \qquad (11)$$

where $\mu = \cos\theta$.

The damage energy is given by:

$$E_a(\mu, E) = E_R(\mu, E)P(E_R(\mu, E)/E_L). \qquad (12)$$

The emission angular-integrated damage cross section is obtained by:

$$\sigma_D(E) = \sigma(E)\int_{-1}^{1} f(\mu, E)\, E_a(\mu, E)\xi(E_a)d\mu \qquad (13)$$

where $\sigma(E)$ is the corresponding cross section. $\xi(E_a)$ is the efficiency of atomic displacement based on the NRT metric, it is unity for the NRT and Eq. (5) for the ARC model. $f(\mu, E)$ is the probability density of angular distribution for the incident energy $E$ versus the cosine of the emission angle in the CM frame $\mu$. $f(\mu, E)$ is conventionally expressed as a sum of Legendre polynomials:

$$f(\mu, E) = \sum_{l=0}^{Lmax} \frac{2l+1}{2} a_l(E)P_l(\mu) \qquad (14)$$

where $P_l$ is the $l$-th Legendre polynomial and $a_l$ is the corresponding Legendre coefficient given in the ENDFs.

The corresponding DPA cross section is computed by:

$$\sigma_{DPA}(E) = \frac{0.8\sigma_D(E)}{2E_d}. \qquad (15)$$

The advantage of using damage cross section rather than direct DPA cross section is that the former is not sensitive to $E_d$ [24] because it depends on $E_d$ only in $[E_d, 2.5E_d]$. Therefore, the value of $E_d$ used in the calculation of damage cross section is not important for subsequent calculation of DPA rate. Since the DPA cross section and the damage cross section are the same with a factor of $2.5E_d$, we do not distinguish the names of DPA cross section and the damage cross section in the following studies which show always the damage cross sections.

2.3. DPA cross section and energy-angle distribution in the laboratory frame

The conservation of momentum in the Lab frame points out:

$$mv_0 = m'v'\cos\varphi + M'v_R\cos\alpha \tag{16}$$

$$m'v'\sin\varphi = M'v_R\sin\alpha \tag{17}$$

By eliminating $\alpha$, the recoil energy of PKA is obtained as:

$$E_R(E, E', \tilde{\mu}) = \frac{1}{M'}[mE - 2\sqrt{mm'EE'}\tilde{\mu} + m'E'] \tag{18}$$

where $\tilde{\mu} = \cos\varphi$. The energy-angle-integrated damage cross section related to a given reaction is obtained by:

$$\sigma_D(E) = \sigma(E) \int_0^\infty \int_{-1}^1 \tilde{f}(E, E', \tilde{\mu}) E_a(E, E', \tilde{\mu}) \xi(E_a) d\tilde{\mu} dE' \tag{19}$$

where $\tilde{f}(E, E', \tilde{\mu})$ is the probability density of energy-angular distribution in the Lab frame for the incident energy $E$ versus the secondary energy $E'$ and the cosine of the emission angle $\tilde{\mu}$. $\tilde{f}(\tilde{\mu}, E, E')$ is conventionally given by the combination of Legendre polynomials:

$$\tilde{f}(E, E', \tilde{\mu}) = \sum_{l=0}^{Lmax} \frac{2l+1}{2} \tilde{b}_l(E, E') P_l(\tilde{\mu}) \tag{20}$$

where $\tilde{b}_l(E, E')$ is the Legendre coefficient given in ENDF.

2.4. DPA cross section and energy-angle distribution in the center of mass frame

Section 2.3 shows the method for the energy-angle distributions are given in the Lab frame in ENDF [25]. However, the double differential cross sections are often provided in the CM frame on which many nuclear theories are based, such as that of $^{56}$Fe of JEFF-3.1.1 [7]. Three methods can be used to compute the DPA cross section with double-differential cross sections given in the CM frame. These methods should be found in some textbooks. However, since it is hard to find a book summarizing all methods for damage calculations, the present work briefly shows the calculations of damage with double-differential cross sections given in the CM frame.

*2.4.1. Relationship between variables in the CM frame and the Lab frame*

**Lab to CM:** The velocity of the emission particle in the CM frame is (see the lower right scheme in Figure 1):

$$u_1^2 = v'^2 + v_{CM}^2 + 2v'v_{CM}\tilde{\mu} \tag{21}$$

The explicit expression of $E_1$ with the quantities in the Lab frame is thus:

$$E_1 = E' + \frac{mm'E}{(m+M)^2} - 2\sqrt{\frac{mm'EE'}{(m+M)^2}}\tilde{\mu} \tag{22}$$

Projecting the velocity into the incident direction leads to:

$$u_1\mu + v_{CM} = v'\tilde{\mu} \tag{23}$$

Thus, the explicit expression of $\mu$ knowing $\tilde{\mu}$ is:

$$\mu = \sqrt{\frac{E'}{E_1}}\left(\tilde{\mu} - \sqrt{\frac{mm'E}{(m+M)^2 E'}}\right) \tag{24}$$

**CM to Lab:** Transforming the recoil velocity of the emission particle from the CM to the Lab frame:

$$v'^2 = u_1^2 + v_{CM}^2 + 2u_1 v_{CM}\mu \tag{25}$$

Consequently, the secondary energy in the Lab frame is:

$$E' = \frac{mm'E}{(m+M)^2} + E_1 + 2\frac{\sqrt{mm'EE_1}}{m+M}\mu \qquad (26)$$

On the other hand, Eq. (23) implies:

$$\tilde{\mu} = \frac{u_1\mu + v_{CM}}{v'} \qquad (27)$$

One can further obtain the expression:

$$\tilde{\mu} = \frac{\sqrt{mm'E} + (m+M)\sqrt{E_1}\mu}{\sqrt{(m+M)^2 E_1 + mm'E + 2(m+M)\sqrt{mm'EE_1}\mu}} \qquad (28)$$

*2.4.2. Transformation of data from the CM frame to the Lab frame*

Section 2.3 shows the routine of DPA calculations with energy-angular distribution in the Lab frame. For the data given in the CM frame, this method can be applied by transforming the data in the CM frame to the Lab frame. The transformation of data from the CM frame to the Lab frame is also the strategy of NJOY [5]. For a given incident energy $E$, the coefficients $\tilde{b}_l(E, E')$ in Eq. (20) can be determined through the energy-angular distribution provided in the CM frame $f(E, E_1, \mu)$. This method is implemented in NJOY because the Legendre coefficients in the Lab frame can be used to compute all corresponding quantities in the same frame.

Because the Legendre polynomials are orthogonal (and orthonormal for $((2l+1)/2)^{1/2}P_l$) with respect to the $L^2$ norm on the interval [-1,1], the coefficients $\tilde{b}_l(E, E')$ are defined by:

$$\tilde{b}_l(E, E') = \int_{-1}^{1} \tilde{f}(E, E', \tilde{\mu}) P_l(\tilde{\mu}) d\tilde{\mu} \qquad (29)$$

Since there are two degrees of freedom for the energy-angular distribution, the transformation from the CM frame to the Lab frame should be performed with a double integral:

$$\int_{-1}^{1} \tilde{f}(E, E', \tilde{\mu}) P_l(\tilde{\mu}) d\tilde{\mu} = \int_{-1}^{1} \int_{0}^{E'_{max}} \delta_{E'}(E'') \tilde{f}(E, E'', \tilde{\mu}) P_l(\tilde{\mu}) dE'' d\tilde{\mu} \qquad (30)$$

where $E'_{max}$ can be determined by Eq. (26) and the Dirac delta function about $E'$ is defined as:

$$\delta_{E'}(E'') = \begin{cases} 1, & E'' = E' \\ 0, & \text{otherwise} \end{cases} \qquad (31)$$

By using the data in the CM frame, the Legendre coefficients in the Lab frame are:

$$\tilde{b}_l(E, E') = \int_{-1}^{1} \int_{0}^{E_{1,max}} \delta_{E'}(E_1, \mu) f(E, E_1, \mu) P_l(\tilde{\mu}(E_1, \mu)) dE_1 d\mu \qquad (32)$$

where the Dirac delta function links two variables in the CM frame with $E'$:

$$\delta_{E'}(E_1, \mu) = \begin{cases} 1, & \text{Eq. (26): } (E_1, \mu) \to E' \\ 0, & \text{otherwise} \end{cases} \qquad (33)$$

The maximum secondary energy in the CM frame $E_{1,max}$ in Eq. (32) is directly given in ENDF.

The change of variables in double integrals for Eq. (32) conducts to:

$$\tilde{b}_l(E, E') = \int_{\tilde{\mu}_{min}}^{1} f(E, E_1(E', \tilde{\mu}), \mu(E', \tilde{\mu})) P_l(\tilde{\mu}) J(E) d\tilde{\mu} \qquad (34)$$

where the Jacobian $J(E)$ is the determinant of the Jacobian matrix of the transformation from $(E', \tilde{\mu})$ to $(E_1, \mu)$:

$$Jac(E, E', \tilde{\mu}) = \begin{bmatrix} \partial E_1/\partial E' & \partial E_1/\partial \tilde{\mu} \\ \partial \mu/\partial E' & \partial \mu/\partial \tilde{\mu} \end{bmatrix} \quad (35)$$

The determinant of the Jacobian matrix calculated with Eqs. (22), (24), (26) and (35) is:

$$J(E) \equiv \det[Jac(E, E_1, \mu)] = \sqrt{\frac{E_1}{E'}} \quad (36)$$

$J(E)$ rather than $|J(E)|$ is used in Eq. (34) because the Jacobian is always positive, shown by Eq. (36).

The lower limit of the integral in Eq. (34) is not necessarily -1 because the minimum value of $\tilde{\mu}$ for a given $E'$ can be larger than -1. This is due to the limits in [-1,1] for $\mu(E', \tilde{\mu})$. According to Eqs. (26) and (28), the lower limit of the integration in Eq. (34) is:

$$\tilde{\mu}_{min}(E_{1,max}) = \max\left\{ \frac{\sqrt{mm'E}}{(m+M)\sqrt{E'}} - \sqrt{\frac{E_{1,max}}{E'}}, \; -1 \right\} \quad (37)$$

For a given $(E, E')$, $E_1$ is a function of $\tilde{\mu}$. Calculation of $\tilde{b}_l(E, E')$ by Eq. (34) requires the density $f(E, E_1(E', \tilde{\mu}), \mu(E', \tilde{\mu}))$ for each $\tilde{\mu}$. The energy-angular distributions are usually tabulated for the secondary energy $E_1$. The interpolation of $f(E, E_1, \mu)$ on the secondary energy grid is required for each $\tilde{\mu}$. This method increases the computation burden.

Moreover, for a given incident energy, we should define a suitable grid of the secondary energy. If the grid is too fine, too many calculations and storages are required. If the grid is too coarse, some information will be lost. NJOY takes the criterion that the difference between the coefficient of the midpoint in each interval calculated by Eq. (34) and the linearly interpolated value with two boundaries should be less than 2% [26]. Anyway, transforming the data of energy-angular distribution in the CM frame to the Lab frame gives an additional error for DPA cross section.

### 2.4.3. Change of variables in double integrals

The change of variables is an intuitive method for the transformation of frames. This method can avoid the problem of the loss of information. Because the Jacobian is always positive, the change of double variables in the CM frame to the Lab frame leads to:

$$\sigma_D(E) = \sigma(E) \int_0^\infty \int_{-1}^1 f(E, E_1, \mu) E_a\big(E, E'(E_1, \mu), \tilde{\mu}(E_1, \mu)\big) \xi\big(E_a(E_1, \mu)\big) [J(E)]^{-1} d\mu dE_1 \quad (38)$$

where the Jacobian $J(E)$ is found in Eq. (36) with $E'$ in Eq. (26), $\tilde{\mu}(E_1, \mu)$ is given in Eq. (28).

The DPA cross sections with the energy-angular distributions provided in the CM frame can be computed with Eq. (38). All information given in the CM frame can be used in the computation of DPA cross sections. However, the integrand in Eq. (38) has a much more complex form than the integrand in Eq. (19). A consequent result is that the numerical integration for Eq. (38) converges more slowly than Eq. (19). In other words, comparing with the energy-angular distribution given in the Lab frame and DPA cross sections computed with Eq. (19), finer grids are required to perform the numerical integrals of Eq. (38) in the case of double-differential nuclear data given in the CM frame.

*2.4.4. Direct calculation in the CM frame*

The above methods can compute the DPA cross sections with the energy-angular distribution in the CM frame. The transformation of data between two frames increases the computation burden and storage memory, and introduces additional error. The change of variables is more feasible than the transformation of frames for the calculations of DPA cross sections. However, compared with the double-differential nuclear data given in the Lab frame and the DPA cross sections calculated by Eq. (19), the integrant in the change of variables method has a more complex form.

In fact, the direct calculation of DPA cross sections in the CM frame is much simpler than the two previous methods for the energy-angular distributions given in the same frame. Using Eqs. (7) and (9), the recoil energy can be obtained as:

$$E_R(E, E_1, \mu) = \frac{mM'}{(m+M)^2}E - 2\frac{\sqrt{mm'EE_1}}{m+M}\mu + \frac{m'}{M'}E_1 \qquad (39)$$

The energy-angle-integrated DPA cross section can be directly computed with:

$$\sigma_D(E) = \sigma(E)\int_0^\infty \int_{-1}^1 f(E, E_1, \mu)\, E_a(E, E_1, \mu)\xi(E_a)d\mu dE_1 \qquad (40)$$

where $f(E, E_1, \mu)$ is the probability density of energy-angular distribution in the CM frame. $f(\mu, E, E_1)$ is conventionally expressed by:

$$f(E, E_1, \mu) = \sum_{l=0}^{Lmax} \frac{2l+1}{2} b_l(E, E_1) P_l(\mu) \qquad (41)$$

where $b_l(E, E_1)$ is the Legendre coefficient in the CM frame given in ENDF.

Comparing with the change of variables, this method simplifies the calculations. In addition, the integrand in Eq. (40) has a simpler form than that in Eq. (38). Therefore, numerical methods converge more quickly for the direct calculation in the CM frame than the change of variables. As a matter of fact, Eqs. (19) and (40) have the similar form, the computation of DPA cross sections with the energy-angular distributions given in the CM frame by Eq. (40) converges as quickly as the calculation with double-differential data provided in the Lab frame by Eq. (19). Therefore, the method of direct computation in the CM frame with Eq. (40) is recommended if the energy-angular distributions are given in the CM frame.

## 3. Refinements of DPA cross section calculations

3.1. Refinement of DPA calculations with angular distributions

The DPA cross sections proposed in Sections 2.2-2.4 are based on the formula $DPA(E_a) = 0.8E_a/2E_d\xi(E_a)$. The latter is available only for $E_a > 2E_d/0.8$ according to the DPA metrics given in Eqs. (2) and (4). In order to use the same expression of DPA in the whole domain, one generalizes the damage energy in the interval $[0, 2E_d/0.8]$ as:

$$\widetilde{E_a}(\mu, E) = \begin{cases} 0, & 0 < E_a < E_d \\ 2E_d/0.8, & E_d < E_a < 2E_d/0.8 \\ E_a(\mu, E), & 2E_d/0.8 < E_a \end{cases} \quad (42)$$

Therefore, $DPA(\widetilde{E_a}) = 0.8\widetilde{E_a}/2E_d \xi(\widetilde{E_a})$ is valid in the whole range. The damage cross sections mentioned in Sections 2.2 to 2.4 are thus available for any physical value of $E$ or $\mu$ by using the generalized damage energy. The computation of DPA cross section is simplified due to the same expression in the whole domain. It is noticeable that the second "stair" is not accounted in the widely used code NJOY [26]. Users should add this interval for the partition function in the HEATR module. To simplify the notation, the generalized damage energy in Eq. (42) is also called as the damage energy hereinafter. Figure 2 illustrates the damage energy of 5 keV neutron elastic scattering on $^{56}$Fe.

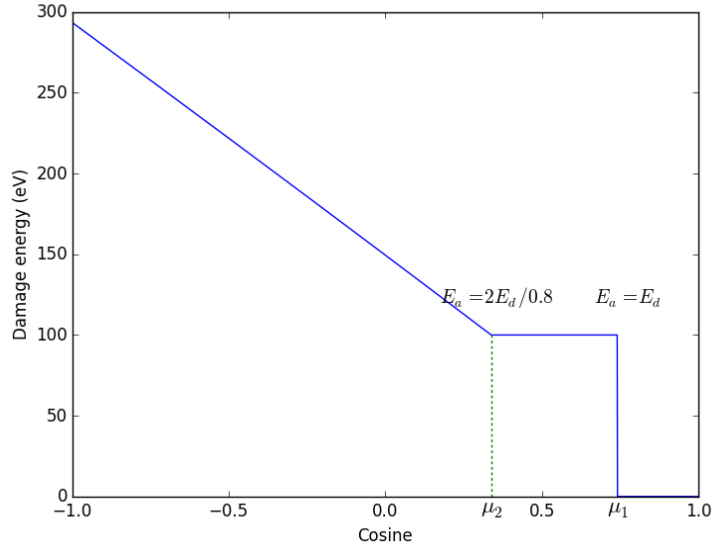

Figure 2. Damage energy of $^{56}$Fe versus the cosine in CM for 5 keV neutron elastic scattering

To calculate the integral versus emission angle, a 64-point Gauss-Legendre Quadrature (GLQ) method is used in NJOY2016 (20-point in the manual) [26]. However, the damage energy is not a continuous function of the cosine of the emission angle (Eq. (42)), so neither the product with the angular distribution is. 64-point GLQ cannot necessarily ensure the accuracy of the integral. As the standard metric, the NRT formula is used in numerical examples in the following studies. Figure 3 indicates the neutron elastic DPA cross sections of $^{56}$Fe computed with different numbers of points in the GLQ. The DPA cross section does not converge for the 150-point GLQ at neutron energy below 10 keV because of the large contribution of damage energy in [0, $2E_d/0.8$]. The integral converges at high incident energy because the damage energy lower than $2E_d/0.8$ is less important. This range is not important for most reactions because of the small angle-integrated damage energy. However, for some reactions having resonances in this range, the DPA cross sections can have large influence on DPA rate calculations because the DPA cross section is the product of reaction cross section and the angle-integrated (energy-angle-integrated for continuum reactions) damage energy.

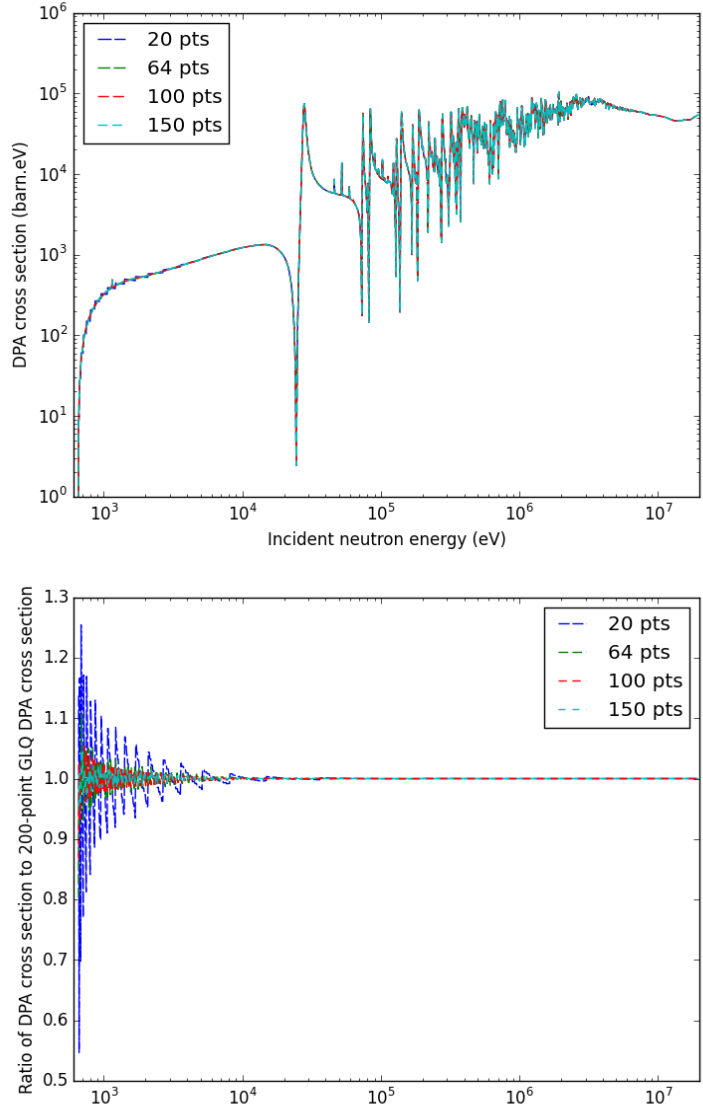

Figure 3. Neutron elastic scattering DPA cross sections of $^{56}$Fe performed with 20, 64, 100, and 150 points Gauss-Legendre quadrature (upper) and the corresponding ratios to the 200-point Gauss-Legendre quadrature calculation (lower).

In order to ensure the convergence for computing the DPA cross sections, the present work proposes to compute the integral in three intervals according to the damage energy. The two critical points to connect the three intervals are obtained with:

$$E_a(\mu_1, E) = E_d \qquad (43)$$
$$E_a(\mu_2, E) = 2E_d/0.8 \qquad (44)$$

With the notations of $\mu_1$ and $\mu_2$, $E_a > 2E_d/0.8$ for $\mu$ in the interval [-1, $\mu_2$]; $\mu$ in [$\mu_2$, $\mu_1$] is equivalent to damage energy in [$E_d, 2E_d/0.8$], so the damage energy is $2E_d/0.8$; for $\mu > \mu_1$, the damage energy is zero. The examples on $\mu_1$ and $\mu_2$ are pointed out in Figure 2.

In fact, by definition of the threshold energy, a PKA with energy higher than $E_d$ should be displaced. In addition, if the kinetic energy of a PKA is lower than $2E_d$ ($2E_d/0.8$ by accounting the efficiency), this PKA cannot induce a second vacancy. The energy transferred to the electronic excitation during and after collision has no influence on DPA in this range. It is better to use the PKA energy rather than the damage energy in the interval $[0, 2E_d/0.8]$. Accordingly, Eqs. (43) and (44) become:

$$E_R(\mu_1, E) = E_d \qquad (45)$$
$$E_R(\mu_2, E) = 2E_d/0.8 \qquad (46)$$

Taking the limits of the cosine into account, the boundaries are:

$$\mu_1(E) = p_{[-1,1]}\left(\left[\frac{mM'}{m'M} + R(E)^2 - \frac{(m+M)^2}{m'ME}E_d\right] / \left[2R(E)\sqrt{\frac{mM'}{m'M}}\right]\right) \qquad (47)$$

$$\mu_2(E) = p_{[-1,1]}\left(\left[\frac{mM'}{m'M} + R(E)^2 - \frac{2(m+M)^2}{0.8m'ME}E_d\right] / \left[2R(E)\sqrt{\frac{mM'}{m'M}}\right]\right) \qquad (48)$$

where $p_{[-1,1]}$ is the projection on [-1,1] defined by:

$$p_{[-1,1]}(x) = \begin{cases} -1, & x < -1 \\ x & -1 \leq x \leq 1 \\ 1, & x > 1 \end{cases} \qquad (49)$$

According to Eq. (42), the NRT-DPA cross sections are computed with:

$$\sigma_D(E) = \sigma(E)\left[\int_{-1}^{\mu_2} f(\mu, E) E_R(\mu, E) P\left(\frac{E_R(\mu, E)}{E_L}\right) d\mu + \frac{2E_d}{0.8}\int_{\mu_2}^{\mu_1} f(\mu, E) d\mu\right] \qquad (50)$$

This GLQ method applied in DPA calculations is referred to the GLQ based Piecewise Integration (GLQPI) hereinafter. The ARC-DPA cross sections are calculated by inserting the efficiency $\xi(E_R(\mu, E) P(E_R(\mu, E)/E_L))$ in the first integral in Eq. (50).

Let Lmax denote the highest order of Legendre polynomials for describing $f(\mu, E)$. Because the *N*-point GLQ compute the exact value of integral for the (2*N*-1)-order polynomials, [Lmax/2+1]-point GLQ is sufficient to accurately compute the second integration in Eq. (50). For elastic and discrete inelastic scattering of JEFF-3.1.1, Lmax = 19 reveals that 10-point GLQ is sufficient to compute the second integration in Eq. (50). Because the integrand in the first integration in Eq. (50) is the product of a (Lmax+1)-order (Lmax for *f* and a cosine in recoil energy) polynomial and a smooth but non-polynomial partition function, more than [(Lmax+1)/2]+1 points (11 points for $^{56}$Fe of JEFF-3.1.1) should be used. Figure 4 indicates the DPA cross sections with 20 points and 200 points GLQPI. The excellent agreement between the DPA cross sections calculated with 20-point GLQPI and 200-point GLQPI points out the convergence of the integral. It is noticeable that the maximum order can be up to 64 in ENDF-6 [25], that signifies more than 33 points are required. For the purpose of verification, more than 33 points should be used as a reference to verify the convergence of numeric integration. However, due to the negligible contribution of high-order Legendre polynomials on damage cross sections (c.f. Section 4), fewer points are required for the numeric integration.

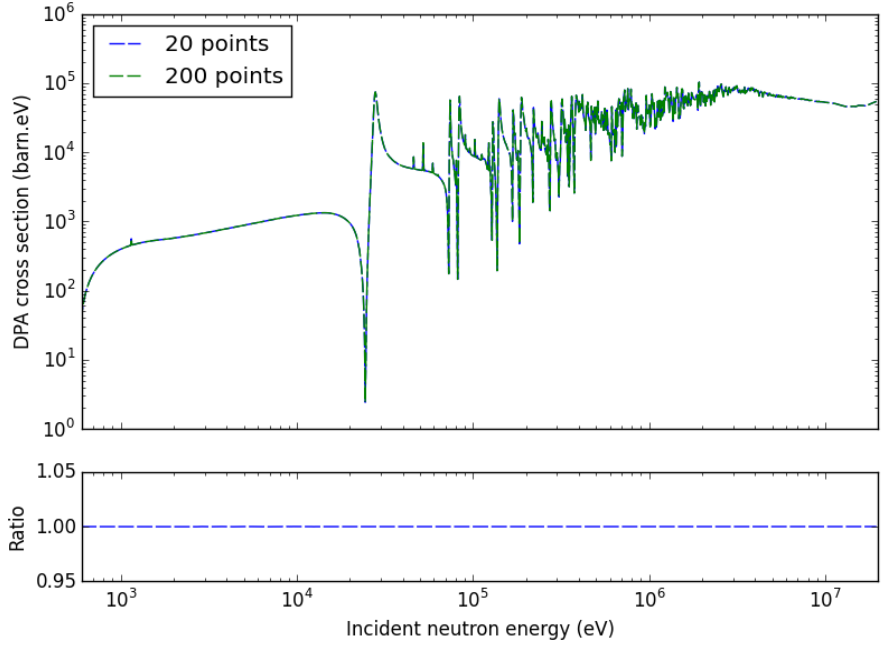

Figure 4. Neutron elastic scattering DPA cross sections of $^{56}$Fe performed with 20 and 200 points Gauss-Legendre Quadrature based Piecewise Integration (GLQPI).

3.2. Integration over the secondary energy

For the elastic and discrete inelastic scatterings, the DPA cross sections are computed with only the angular integration because the secondary energy is determined by the emission angle and the constant $Q$-value of the reaction. In other words, the elastic and discrete inelastic scatterings have only 1 degree of freedom. For the continuum inelastic scatterings, the integration over the secondary energy is required because the emission angle and the secondary energy are two independent variables. The secondary energy distribution of the continuum inelastic scattering of $^{56}$Fe is plotted in Figure 5. Because the angular integrations are performed with the GLQ,

$$\int_0^\infty \int_{-1}^1 f(E, E_1, \mu) \, E_a(E, E_1, \mu) d\mu dE_1 = \sum_{i=1}^N w_i \int_0^\infty f(E, E_1, x_i) E_a(E, E_1, x_i) \, dE_1 \quad (51)$$

where $w_i$ and $x_i$ are respectively the $i$-th weight and the $i$-th Gauss node ($i$-th zero of $P_N$) in the $N$-point GLQ.

The probability density function $f(E, E_1, x_i)$ is conventionally a linearly interpolated function between two secondary energies. The recoil PKA energy is not a polynomial function as the secondary energy due to the square root term (shown in Eq. (39)). The partition function is never a polynomial function of the secondary energy. The integrant in Eq. (51) is thus non polynomial. For the tabulated secondary energy distribution, which is often the case, it is reasonable to use the trapezoidal integration with the energy grid in the ENDF. Because the error of the trapezoidal integration on each interval is dominated by the second derivative of the integrand, the accuracy depends on the energy grid.

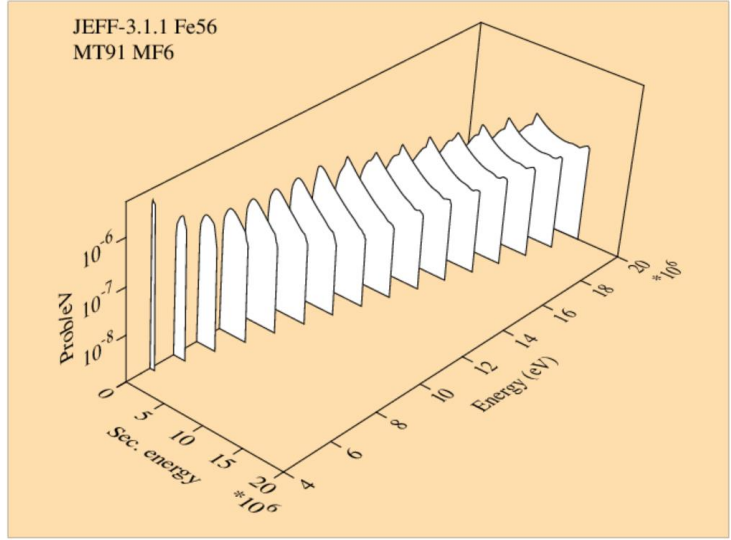

Figure 5. Energy distribution of the continuum inelastic scattering of $^{56}$Fe in JEFF-3.1.1 plotted by NJOY-2016

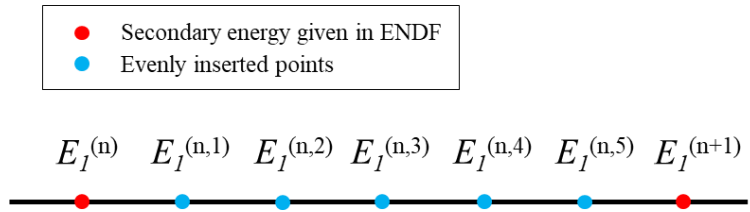

Figure 6. Scheme of 5 evenly inserted points between two neighbor secondary energies given in ENDF

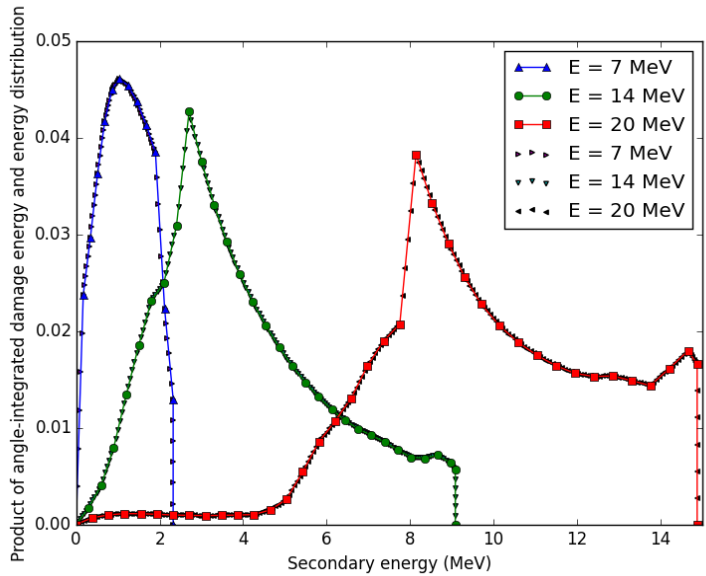

Figure 7. Energy distribution of the angle-integrated damage energy, i.e. $\int_{-1}^{1} f(E, E_1, \mu) \, E_a(E, E_1, \mu) d\mu$, for 7 MeV, 14 MeV, and 20 MeV neutron continuum inelastic scattering of $^{56}$Fe. The larger points are calculated based on the energy grid in JEFF-3.1.1. The smaller points are computed at 5 evenly inserted energies.

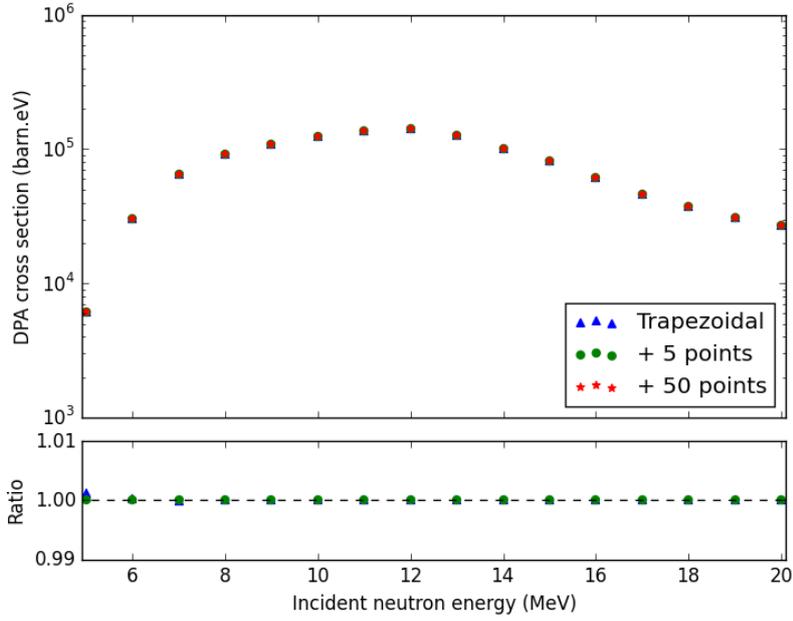

Figure 8. Neutron continuum inelastic scattering DPA cross sections of $^{56}$Fe computed with the trapezoidal integration by using the energy grid in ENDF, 5 and 50 evenly inserted points in each interval. The corresponding ratios are calculated over the DPA with 50 inserted points.

To verify the convergence of the trapezoidal integration in the calculations of DPA cross sections, the present work proposes to evenly add points in each interval of the original energy grid for numeric integral by interpolating the energy distributions. The example of the 5 evenly inserted points illustrated in Figure 6. The energy distributions of the angle-integrated damage energy of $^{56}$Fe are shown in Figure 7 for 7 MeV, 14 MeV, and 20 MeV incident energies. The larger scattered points are computed with the data in JEFF-3.1.1. The lines are the linear interpolation of scattered points. The calculations performed with the 5 evenly inserted points are shown in Figure 7 with smaller triangles. The results of direct interpolation are quite similar to the interpolated nuclear data-based calculations. Figure 8 shows the DPA cross sections of the continuum inelastic scattering of $^{56}$Fe performed with the trapezoidal integration using the energy grid in JEFF-3.1.1, 5- and 50-inserted equidistant points in each interval of the secondary energy in JEFF-3.1.1. The numerical results show that the trapezoidal integration using the energy grid in ENDF can give accurate results for $^{56}$Fe, while 2% potential error is permitted in NJOY during the transformation of data from the CM frame to the Lab frame [5]. However, because the accuracy of the numeric integration depends on the grid, the verification with the above method is always recommended to ensure the accuracy of DPA cross section for each continuum reaction.

3.3. Computation of DPA cross sections between two incident energies

The above analyses of DPA cross sections for continuum reactions are based on the incident energy at which the energy-angular distribution is given in the ENDF. To compute the DPA cross sections between two neighbor incident energies, NJOY uses the linear interpolation of the energy-angle-integrated damage energy. In fact, the physical method is to compute DPA cross sections using the interpolated the energy-angular distribution at each energy, as the example shown in Figure 9 with green points. The most common method for interpolating energy-angular distributions is the Unit-Base Interpolation (UBI) [27]. For linear-linear UBI, knowing the energy distribution at two incident energies $E_{n,0}$ and $E_{n,1}$, the probability for incident energy of $E$ and secondary energy $E'$ is given by [27]:

$$P(E, E') = \frac{\tilde{P}(E,\widetilde{E'})}{E'_{max,q}-E'_{min,q}} \tag{52}$$

where

$$q = \frac{E-E_{n,0}}{E_{n,1}-E_{n,0}} \tag{53}$$

$$E'_{max/min,q} = (1-q)E'_{max/min,0} + qE'_{max/min,1} \tag{54}$$

$$\widetilde{E'} = \frac{E'-E'_{min,q}}{E'_{max,q}-E'_{min,q}} \tag{55}$$

$$\tilde{P}(E,\widetilde{E'}) = (1-q)\tilde{P}(E_{n,0},\widetilde{E'}) + q\tilde{P}(E_{n,1},\widetilde{E'}) \tag{56}$$

$$\tilde{P}(E_{n,0/1},\widetilde{E'}) = (E'_{max,0/1} - E'_{min,0/1})P(E_{n,0/1},\widetilde{E'}(E'_{max,0/1} - E'_{min,0/1}) + E'_{min,0/1}) \tag{57}$$

where $P(E_{n,0/1},\widetilde{E'}(E'_{max,0/1} - E'_{min,0/1}) + E'_{min,0/1})$ is given in ENDF. Figure 10 shows the energy distribution for incident energies between 19 MeV and 20 MeV for continuum inelastic scattering of $^{56}$Fe. The right figure is plotted with the normalized secondary energy $\widetilde{E'}$ and the corresponding probability density $\tilde{P}(E,\widetilde{E'})$.

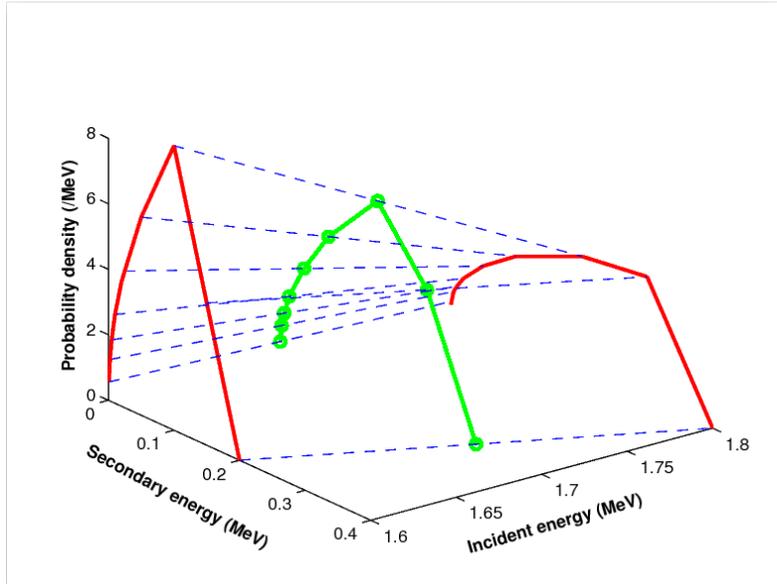

Figure 9. Scheme of the interpolation of energy-angular distributions. Red lines represent the data given in ENDF, the green points are interpolated data.

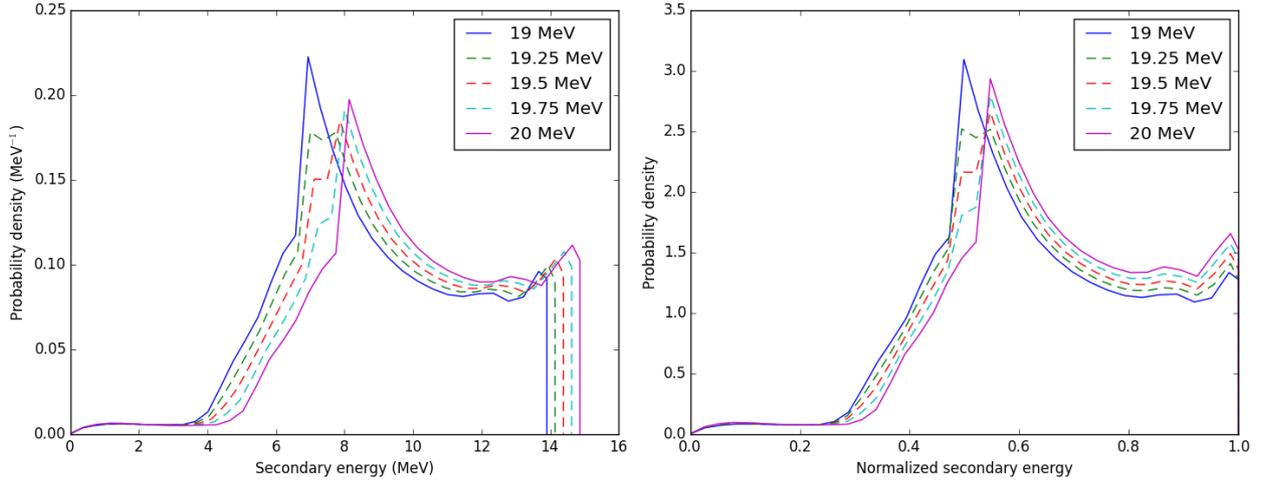

Figure 10. UBI of angle-integrated energy distributions of incident energy between 19 and 20 MeV neutron continuum inelastic scattering with $^{56}$Fe. The right figure uses the normalized secondary energy to intuitively show the peak values.

As shown in Figure 10, the UBI cannot give reasonable peak value of the energy distribution between two given points. Therefore, the present work proposes a Peak value-based UBI (PUBI) for interpolating the energy distributions. In the PUBI, we divide the secondary energies into two intervals according to the peak values. Then the UBI is used to each interval. Assuming the probability density has an unique global maximum, $E_{m,0/1}$ represents the secondary energy corresponding to the maximum probability density of energy distribution:

$$P(E_{n,0/1}, E_{m,0/1}) = \max_{E'_{0/1}}\{P(E_{n,0/1}, E'_{0/1})\} \tag{58}$$

We suppose that the maximum probability for incident energy $E$ is determined by:

$$E_{m,q}/(E'_{\max,q} - E'_{\min,q}) = (1-q)E_{m,0}/(E'_{\max,0} - E'_{\min,0}) + qE_{m,1}/(E'_{\max,1} - E'_{\min,1}) \tag{59}$$

Let denote:

$$\hat{E} = \begin{cases} \dfrac{E' - E'_{\min,q}}{E_{m,q} - E'_{\min,q}} & E' \leq E_{m,q} \\ \dfrac{E' - E_{m,q}}{E'_{\max,q} - E_{m,q}} & E' > E_{m,q} \end{cases} \tag{60}$$

The energy distribution is expressed by:

$$P(E, E') = \begin{cases} \dfrac{\hat{P}_{\mathrm{I}}(E, \hat{E})}{E_{m,q} - E'_{\min,q}} & E' \leq E_{m,q} \\ \dfrac{\hat{P}_{\mathrm{II}}(E, \hat{E})}{E'_{\max,q} - E_{m,q}} & E' > E_{m,q} \end{cases} \tag{61}$$

where

$$\hat{P}_{\mathrm{I/II}}(E, \hat{E}) = (1-q)\hat{P}_{0,\mathrm{I/II}}(E_{n,0}, \hat{E}) + q\hat{P}_{1,\mathrm{I/II}}(E_{n,1}, \hat{E}) \tag{62}$$

where

$$\begin{cases} \hat{P}_{0/1,\mathrm{I}}(E_{n,0/1}, \hat{E}) = (E_{m,0/1} - E'_{\min,0/1})P(E_{n,0/1}, \hat{E}(E_{m,0/1} - E'_{\min,0/1}) + E'_{\min,0/1}) \\ \hat{P}_{0/1,\mathrm{II}}(E_{n,0/1}, \hat{E}) = (E'_{\max,0/1} - E_{m,0/1})P(E_{n,0/1}, \hat{E}(E'_{\max,0/1} - E_{m,0/1}) + E_{m,0/1}) \end{cases} \quad (63)$$

The results corresponding to Figure 10 but with PUBI method are shown in Figure 11. The peak values and the corresponding secondary energies are monotonic for the data obtained by the PUBI method. Figure 11 shows physically reasonable energy distributions for incident energies between two given neighbor energies.

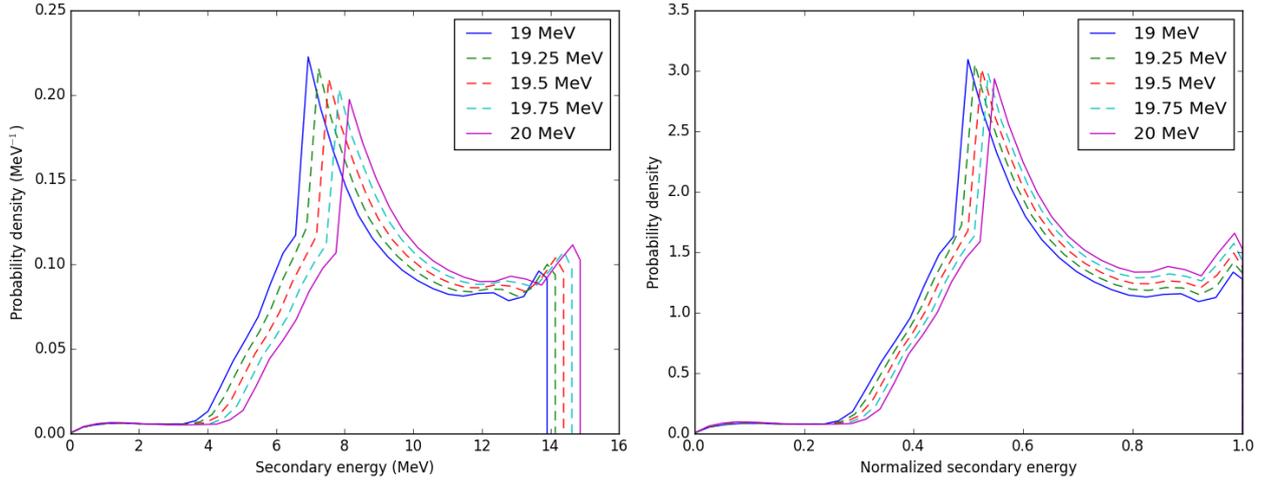

Figure 11. Same results as Figure 10 but with PUBI

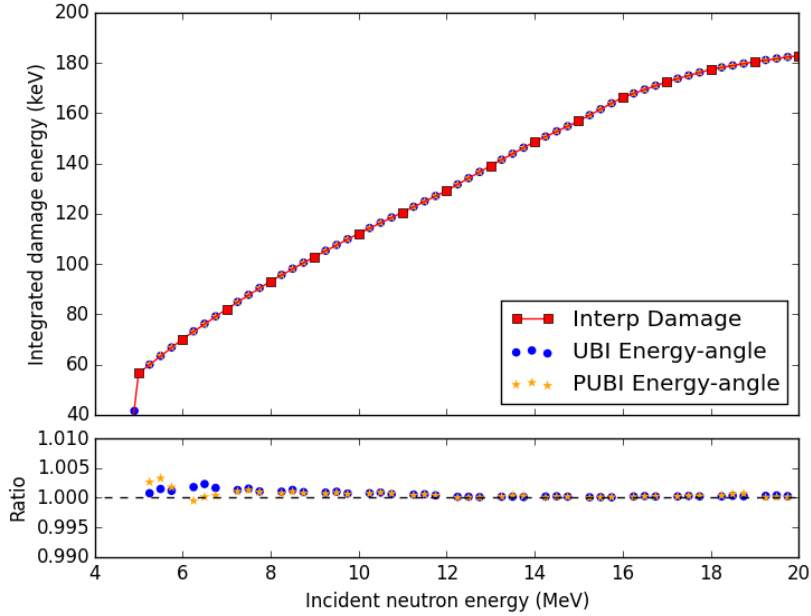

Figure 12. Energy-angle-integrated damage energy. The red square points are computed with energy-angular distributions given in the ENDF, the red line is linear interpolation between two incident energies, the blue circles (orange stars) are calculated with UBI (PUBI) energy-angular distributions. The lower figure plots the ratio to the directly interpolated damage energies.

In order to verify the direct interpolation of energy-angle-integrated damage energy, the DPA cross sections computed with the interpolation of damage energy are compared with those calculated with interpolated energy-angular distributions. Figure 12 reveals the energy-angle-integrated damage energies at different incident energies. The red square points are computed with energy-angular distributions given in ENDF, the red line is linear interpolation between two incident energies, the blue circles (orange stars) are calculated with UBI (PUBI) energy-angular distributions. The ratios of damage energies computed with interpolated energy-angular distributions to the linear interpolated damage energies are shown in the lower sub-plot. The discrepancies are within 0.5%. Therefore, the direct interpolation of energy-angle-integrated damage energies can give the same results as the damage computed with energy-angular distributions computed by standard and improved interpolations. By consequence, the computation of DPA cross sections can be largely simplified using the direct interpolation of energy-angle-integrated damage energies. Again, for a coarse incident grid, the verifications are always recommended to ensure the accuracy of DPA cross sections.

## 4. DPA and high-order Legendre polynomials

4.1. DPA cross sections and Legendre orders

The angular distribution is conventionally presented in the form of Legendre polynomials of the emission angle. The high-order Legendre polynomials play an important role in describing the anisotropy of angular distribution. Taking the neutron elastic scattering of $^{56}$Fe in JEFF-3.1.1 [7] as an example, as shown in Figure 13, up to 19$^{th}$ order Legendre polynomials are required to describe the angular distribution for incident energies higher than 17.5 MeV, while only up to 4$^{th}$ order are sufficient at incident energies below 1.4 MeV. As discussed in Section 3, higher order Legendre polynomials requires more points for GLQ to ensure the convergence. In addition, because the Legendre coefficients are strongly correlated [28], more Legendre polynomials need much more calculations for uncertainty propagation due to the much larger size of the covariance matrix. Jouanne showed that the first order Legendre polynomial of $^{56}$Fe is sufficient to determine the neutron fluence with energy higher than 40 keV on the capsule in a Pressurized Water Reactor (PWR) [10]. If high-order Legendre polynomials are not important for DPA calculation, fewer points should be used for the numerical integral. Moreover, the uncertainty propagation from nuclear data to damage cross sections can be largely simplified. It is of interest to investigate the influence of high-order Legendre polynomials on DPA calculations.

Figure 14 shows the neutron elastic scattering DPA cross sections of $^{56}$Fe computed with different maximum Legendre (Lmax) polynomials. Lmax = 0 is equivalent to the isotropic angular distribution. As shown in Figure 13, at high incident energies, the forward-oriented distribution of the emission neutron is more probable than other directions. Eq. (11) indicates the decrease of the recoil energy with the cosine in the CM frame. Therefore, the anisotropic angular distribution has a negative contribution on DPA cross sections. This result is in accordance with the DPA cross sections of $^{58}$Ni studied in Refs. [8], [9]. The first order Legendre polynomial (L1) can describe somewhat the forward-oriented anisotropy. Nevertheless, L1 is not sufficient to reveal the anisotropy for high incident energies. For example, at an incident neutron energy of 20 MeV, L1 contributes 1.23 (i.e. 6.15%) to the value of probability density $f$ at $\mu = 1$, while the probability density is $f(20\,\text{MeV}, \mu = 1) = 20$. In other words, the probability density calculated with L0 and L1 at $\mu = 1$ is only 11% of the value in JEFF-3.1.1. The example shown in Figure 14 points out that Lmax = 3 (Lmax = 4 resp.) has lower than 5% (2% resp.) overestimation for the neutron elastic scattering DPA cross section of $^{56}$Fe with incident energy below 20 MeV.

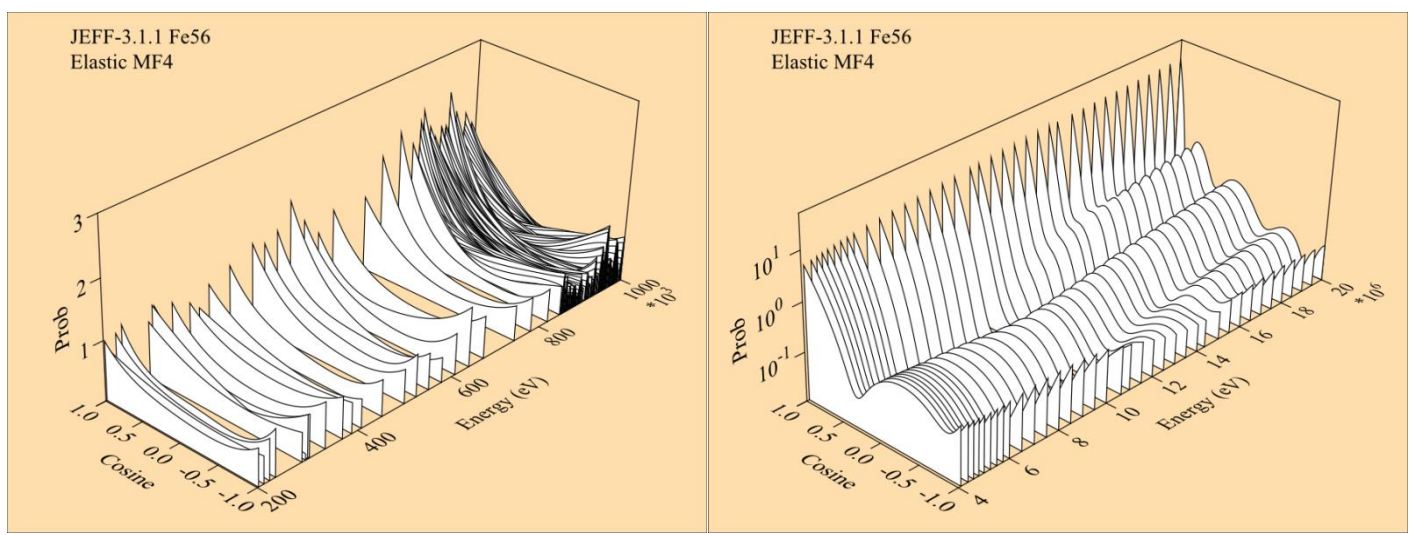

Figure 13. Angular distributions of the neutron elastic scattering reaction of $^{56}$Fe in JEFF-3.1.1 [7] with the incident energies in the interval [200 keV, 1 MeV] (left) and [4 MeV, 20 MeV] (right) plotted by NJOY-2016 [26].

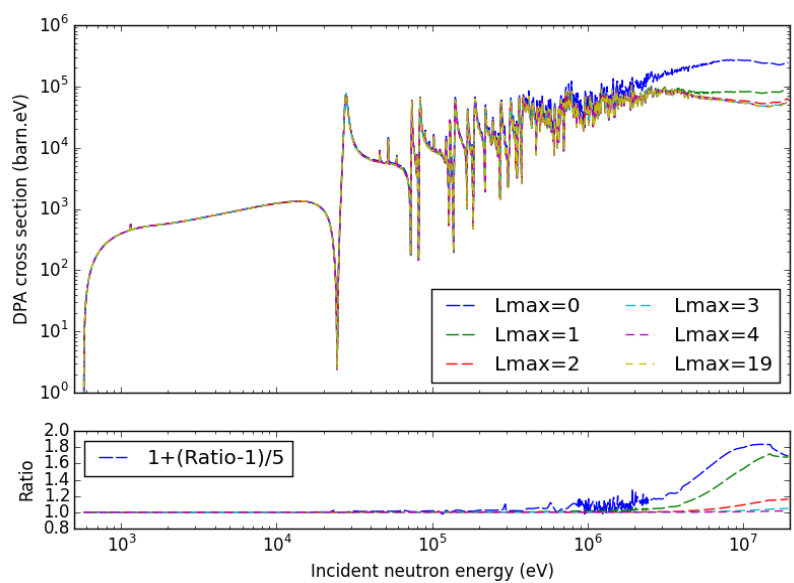

Figure 14. Neutron elastic scattering DPA cross sections of $^{56}$Fe performed with different maximum Legendre (Lmax) polynomials.

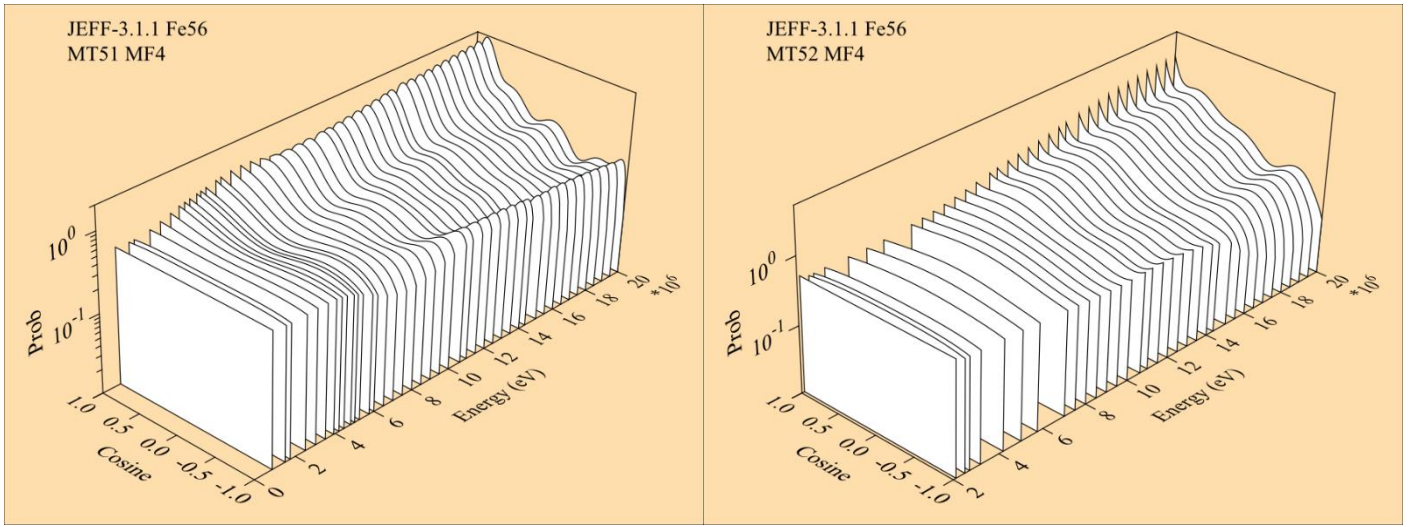

Figure 15. Angular distributions of the neutron first (left) and second (right) levels inelastic scattering reactions (MT51 and MT52) of $^{56}$Fe in JEFF-3.1.1 [7]

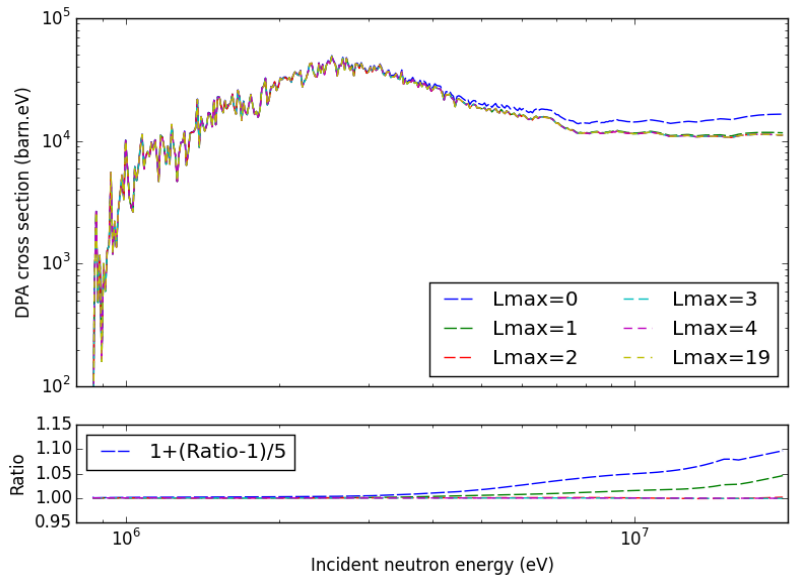

Figure 16. First level neutron inelastic scattering DPA cross sections of $^{56}$Fe performed with different maximum Legendre (Lmax) polynomials.

Compared with elastic scattering, less Legendre polynomials should be used for inelastic scatterings because of the more isotropic angular distribution, as the examples shown in Figure 15. The elastic scattering is more forward-oriented than the inelastic scatterings due to the contribution of the potential scattering. Figure 16 shows the example of the first-level inelastic scattering DPA cross sections. The first order Legendre polynomial can provide enough information on the calculations of DPA cross sections. Since the conclusion is similar for different excitation levels, the results for higher levels of inelastic scattering are not shown in this paper.

Figure 17 illustrates the neutron continuum inelastic scattering DPA cross sections of $^{56}$Fe computed with different Lmax. Figure 17 shows that L1 can well reproduce the DPA cross section at incident energy below 20 MeV, while up to 13$^{th}$ order Legendre polynomials are required to describe the anisotropic angular distributions for the continuum inelastic scattering of $^{56}$Fe in JEFF-3.1.1. Comparing the DPA cross section calculated with full-order Legendre polynomials, the maximum deviation of the DPA cross section calculated with [L0 (isotropic) + L1] is within 1% for incident energy below 20 MeV.

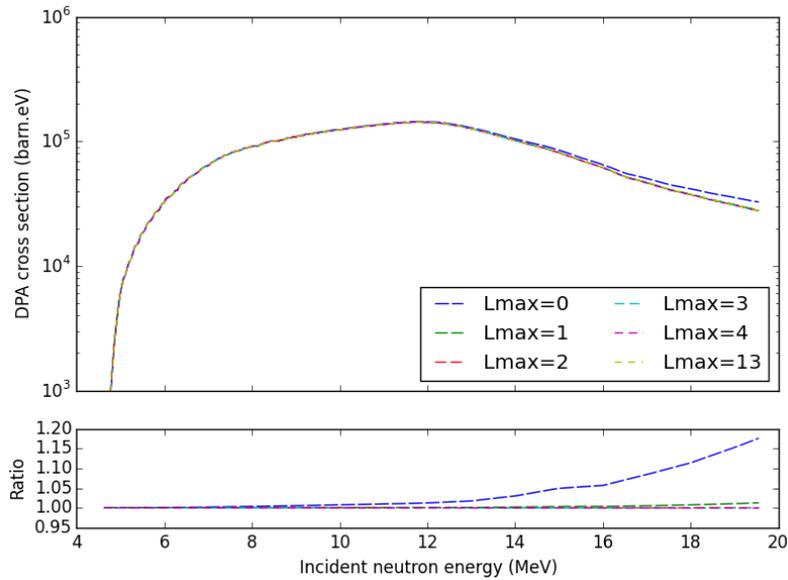

Figure 17. Neutron continuum inelastic scattering DPA cross sections of $^{56}$Fe performed with different maximum Legendre (Lmax) polynomials and the corresponding ratios to the DPA with full order Legendre polynomials.

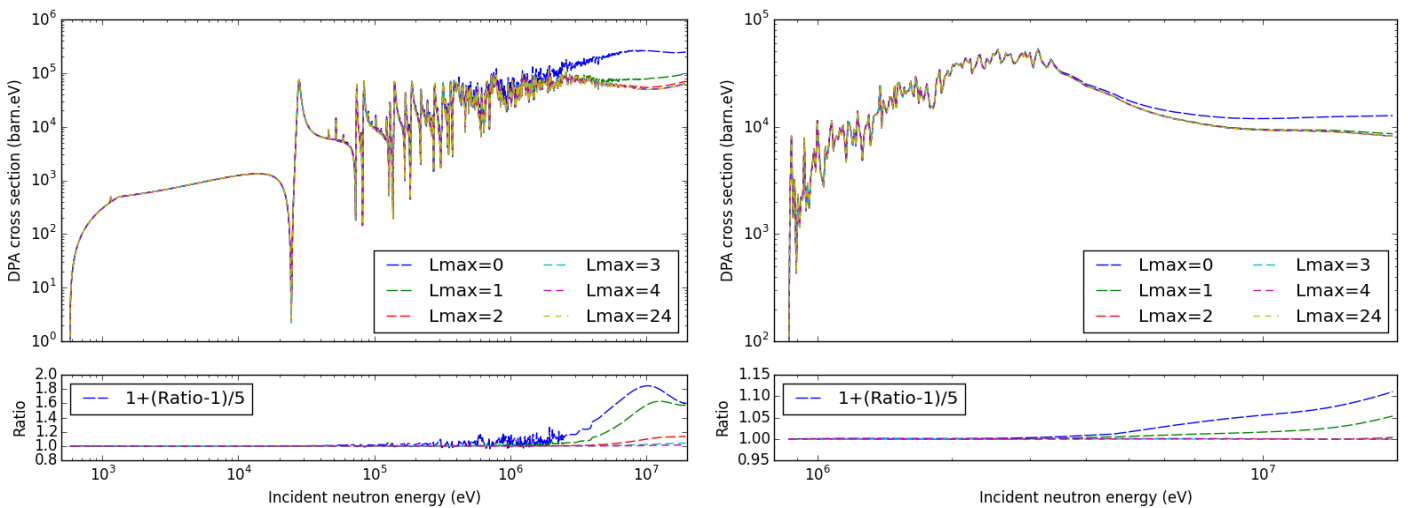

Figure 18. Neutron elastic scattering (left) and first-level inelastic scattering (right)-induced DPA cross sections of $^{56}$Fe with different Lmax using ENDF/B-VIII

Since the above results are based on $^{56}$Fe of JEFF-3.1.1, Figure 18 shows the examples on elastic and first-level inelastic scattering for $^{56}$Fe of ENDF/B-VIII [29]. The ratios to the reference case (i.e. using full-order Legendre polynomial) of $^{56}$Fe are quite similar between JEFF-3.1.1 and ENDF/B-VIII. Even if higher orders of Legendre polynomials are used in $^{56}$Fe of ENDF/B-VIII, the coefficients with order higher than 4 have limited influence on DPA cross sections.

The analysis on the role of high-order Legendre polynomials is also carried out for $^{52}$Cr and $^{58}$Ni, which are the most abundant isotopes in nature Cr and Ni, respectively. Figure 19 illustrates the damage cross section of neutron elastic scattering of $^{58}$Ni using JEFF-3.1.1 and ENDF/B-VIII. The results for $^{52}$Cr and $^{58}$Ni are similar to those of $^{56}$Fe. Therefore, high-order (normally > 4) Legendre polynomials have quite limited influence on DPA cross sections.

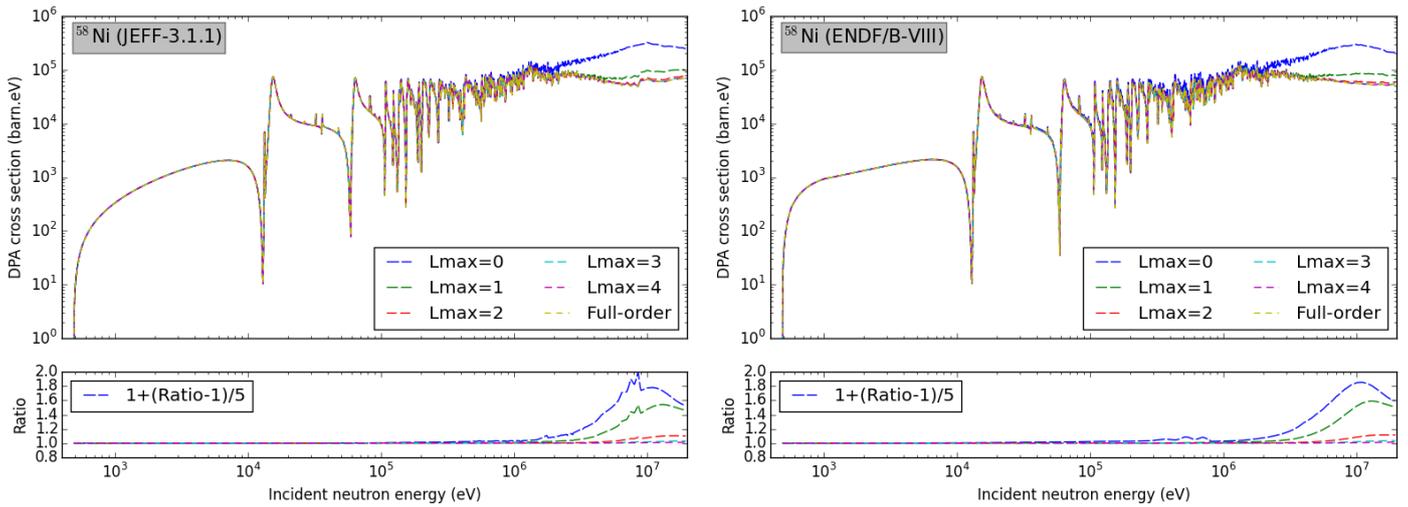

Figure 19. Neutron elastic scattering-induced DPA cross sections of $^{58}$Ni with different Lmax using JEFF-3.1.1 (left) and ENDF/B-VIII (right)

4.2. Application in nuclear facilities

Section 4.1 reveals that the high-order Legendre polynomials are not necessary for DPA cross sections computation because the latter are the angle-integrated values. In order to evaluate the corresponding effect on DPA rates, the examples of the fuel cladding in the Sodium Fast Reactor (SFR) Phenix and the inner surface of RPV in a French 900 MWe PWR are shown because the DPA of the fuel cladding (RPV) in fast reactors (thermal reactors) is the most important factor concerning the fuel cycle length (operating lifetime) reactors. In addition, the DPA rates in a first mirror unit of the Equatorial Visible Infra-Red Wide Angle Viewing System, which is in the diagnostic first wall of the International Thermonuclear Experimental Reactor (ITER) [30], are investigated to evaluate the influence on fusion reactors.

The nuclear data library JEFF-3.1.1 [7] is used to determine the neutron flux. The neutron flux in the fuel cladding of Phenix is calculated with ERANOS [31]. Due to the penetration in the iron, the neutron flux decreases with the depth in RPV. The maximum DPA in RPV is found in the inner surface. The neutron spectrum in the inner surface of RPV in a typical PWR is computed with the stochastic code TRIPOLI-4® [32]. The spectrum of the diagnostic mirror in ITER is also computed by TRIPOLI-4®. The corresponding neutron spectra are shown in Figure 20.

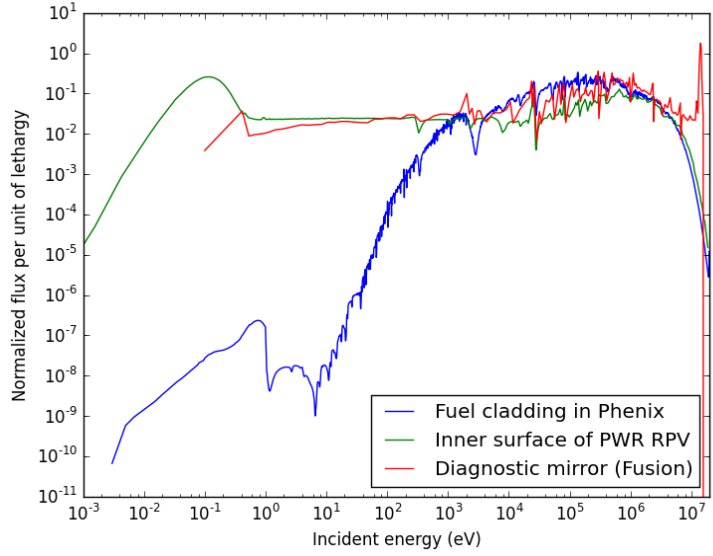

Figure 20. Neutron spectra in the fuel cladding of Phenix reactor, at the inner surface of RPV in a typical PWR, and in the diagnostic mirror of ITER.

Table I. Ratio of DPA rate for $^{56}$Fe with different maximum orders of Legendre polynomials (Lmax) to the reference calculations in different facilities using JEFF-3.1.1 and ENDF/B-VIII (in parenthesis)

|  | Lmax | 0 | 1 | 2 | 3 | 4 | Ref.[a] |
|---|---|---|---|---|---|---|---|
| Phenix | MT2 | 1.405 (1.339)[b] | 1.022 (1.020) | 1.002 | 1.000 | 1.000 | 19.61 (20.09) |
|  | MT51 | 1.022 (1.014) | 1.001 (1.000) | 1.000 | 1.000 | 1.000 | 2.715 (2.922) |
|  | MT52 | 1.007 (1.005) | 0.998 | 1.000 | 1.000 | 1.000 | 0.202 (0.215) |
|  | MT91 | 1.003 | 1.000 | 1.000 | 1.000 | 1.000 | 0.262 (0.262) |
| PWR | MT2 | 1.577 (1.504) | 1.036 (1.034) | 1.004 | 1.001 | 1.000 | 2.10E-3 (2.14E-3) |
|  | MT51 | 1.025 (1.017) | 1.001 (1.000) | 1.000 | 1.000 | 1.000 | 4.23E-4 (4.52E-4) |
|  | MT52 | 1.009 (1.006) | 0.998 | 1.000 | 1.000 | 1.000 | 3.34E-5 (3.46E-5) |
|  | MT91 | 1.004 | 1.000 | 1.000 | 1.000 | 1.000 | 8.08E-5 (8.21E-5) |
| ITER | MT2 | 2.425 (2.309) | 1.192 (1.182) | 1.036 (1.034) | 1.009 | 1.003 | 1.028 (1.071) |
|  | MT51 | 1.125 (1.106) | 1.007 | 1.000 | 1.000 | 1.000 | 0.1917 (0.1899) |
|  | MT52 | 1.083 (1.015) | 1.003 (0.997) | 1.000 | 1.000 | 1.000 | 0.0152 (0.0157) |
|  | MT91 | 1.029 (1.031) | 1.002 (1.001) | 1.000 | 1.000 | 1.000 | 0.5788 (0.5711) |

[a] Reference DPA rate (in DPA/year) with full-order Legendre polynomials
[b] Data in parenthesis are calculated with ENDF/B-VIII but different to the JEFF-3.1.1-based results. No data in parenthesis signifies the same results as JEFF-3.1.1.

The DPA rate induced by particles with a continuous spectrum is calculated by [6]:

$$DPA = \frac{0.8}{2E_d} \int_0^\infty \sigma_D(E)\phi(E)dE \qquad (64)$$

where $\phi(E)$ is the flux of the incident particle. Table I lists the relative DPA rates of $^{56}$Fe with different Lmax in the fuel cladding of Phenix, in the inner surface of RPV in PWR, and just after the first-wall of ITER fusion reactor. MT2, MT51, MT52, and MT91 refer to the elastic, the first-level, the second-level, and the continuum inelastic scatterings, respectively. The DPA rates are normalized by those calculated with all orders of Legendre polynomials provided in ENDF (19 for JEFF-3.1.1 and 24 for ENDF/B-VIII). It is noteworthy that the relative DPA rates for other discrete levels of inelastic scattering are quite similar to MT51 and MT52. The absolute DPA rates of different reactions in different facilities are shown in the last column. In the case that the results based on ENDF/B-VIII are different to those from JEFF-3.1.1, the former data are given in parenthesis.

The numerical results indicate that Lmax = 2 for the elastic scattering can ensure the DPA rates within 0.5% overestimation for fission reactors, while the uncertainties of the neutron elastic scattering cross sections are about 5% [28]. For fusion reactors, up to $4^{th}$ order Legendre polynomials are required to calculate neutron elastic scattering DPA rate. Due to the more isotropic angular distribution as explained in Section 4.1, only the first order Legendre polynomial is necessary to ensure the DPA rates of the inelastic scatterings for both fission and fusion reactors. Because high-order Legendre coefficients and reaction cross sections (and also low-order Legendre coefficient) are strongly correlated [28], the uncertainty quantification in DPA calculations can be largely simplified due to the negligible contribution of high-order Legendre polynomials.

Since the neutron elastic scattering requires higher order of Legendre polynomials, Table II summarizes the neutron elastic scattering-induced DPA rates for $^{58}$Ni and $^{52}$Cr using JEFF-3.1.1 and ENDF/B-VIII. The conclusions about the importance of high-order Legendre polynomials on DPA calculation for $^{58}$Ni and $^{52}$Cr are quite similar to $^{56}$Fe: up to L2 are sufficient for damage calculation in fission reactors, while up to L4 (L1 resp.) for elastic (inelastic resp.) scattering can give DPA rate for fusion reaction within 1% overestimation.

Table II. Ratio of DPA rate for $^{58}$Ni and $^{52}$Cr neutron elastic scattering with different Lmax to the reference calculations using JEFF-3.1.1 and ENDF/B-VIII (in parenthesis)

|  | Lmax | 0 | 1 | 2 | 3 | 4 | Ref.[a] |
|---|---|---|---|---|---|---|---|
| $^{58}$Ni | Phenix | 1.234 (1.298)[b] | 1.013 (1.014) | 1.001 | 1.000 | 1.000 | 35.00 (33.18) |
|  | PWR | 1.384 (1.421) | 1.025 (1.026) | 1.003 | 1.000 (1.001) | 1.000 | .0034 (.0033) |
|  | ITER | 2.080 (2.132) | 1.140 (1.142) | 1.025 | 1.006 | 1.002 | 1.696 (1.593) |
| $^{52}$Cr | Phenix | 1.321 (1.330) | 1.022 | 1.002 | 1.000 | 1.000 | 24.55 (23.23) |
|  | PWR | 1.464 (1.473) | 1.036 (1.037) | 1.004 | 1.001 | 1.000 | .0027 (.0025) |
|  | ITER | 2.103 (3.136) | 1.168 (1.174) | 1.033 (1.035) | 1.008 (1.009) | 1.002 | 1.234 (1.181) |

[a] Reference DPA rate (in DPA/year) with full order Legendre polynomials
[b] Data in parenthesis are calculated with ENDF/B-VIII but different to the JEFF-3.1.1-based results. No data in parenthesis signifies the same results as JEFF-3.1.1.

# 5. Conclusions

The continuum inelastic scattering DPA cross section is computed with the double integral of the energy-angular distribution, which is recommended to be given in the Laboratory (Lab) frame. However, the double-differential cross sections are often given in the Center-of-Mass (CM) frame. Three methods can be applied to perform the DPA calculations with data provided in the CM frame, including the transformation of data from the CM to the Lab frames (used in NJOY), the change of variables in double integrals, and the direct calculation. The first method increases the computation burden because the interpolation is required for each emission angle. Moreover, additional error is introduced due to the loss of information during the transformation. The second one avoids the interpolations and additional error but has a more complex integrant due to the Jacobian. The last method is proposed and recommended in the present work for the energy-angular distributions provided in the CM frame. The direct calculation with the double-differential data in the CM frame is as simple as the method of DPA calculations with the data given in the Lab frame.

The DPA cross sections are computed by integration over the emission angle. The usage of Gauss-Legendre Quadrature (GLQ) in the full range of [-1,1] is shown not convergent for the DPA cross sections due to the discontinuity of the damage energy versus the emission angle. The GLQ-based Piecewise Integration (GLQPI) method is proposed in the present work to ensure the convergence of numerical calculations. The GLQPI method uses the GLQ in each piecewise interval, on which the integrand is a smooth function. For $^{56}$Fe of JEFF-3.1.1, the convergence of the integral is ensured by the 20-point GLQPI, while the 150-point GLQ calculation does not converge.

For continuum reactions, the additional integration over secondary energy is required for DPA cross sections. It is performed with the trapezoidal integration with the energy grid given in ENDF. Because the integrand of the integration over the secondary energy is not a linear function, the method of computations with additional points is proposed to verify the convergence of integration. The comparisons with 5 and 50 evenly inserted points show that the JEFF-3.1.1 energy grid-based trapezoidal integration can calculate accurately the DPA cross sections for $^{56}$Fe.

Because the energy-angular distributions are only tabulated for several incident energies, the DPA cross sections between two neighbor incident energies cannot be directly computed with the double integration over the energy-angular distribution. The present work computes the energy-angular distributions between two incident energies using the standard ENDF interpolation method and an improved method proposed in this paper. The numerical results show that the directly interpolated energy-angle-integrated damage energies correspond well with those computed with the interpolated energy-angular distributions. By consequence, the direct interpolation of energy-angle-integrated damage energies can be used to calculate DPA cross sections.

The high-order Legendre polynomials are of importance to describe the anisotropy of angular distributions. Nevertheless, the high-order Legendre polynomials are shown not important for DPA calculations because the DPA cross section is an angle-integrated quantity. The DPA cross sections computed with isotropic angular distribution are higher than those calculated with the anisotropic emission angle due to the forward-oriented angular distributions, while the damage energy decreases with the cosine of the emission angle. Comparing with inelastic scatterings, higher maximum order of Legendre polynomials is required for the elastic scattering because of the more forward-oriented angular distribution due to the contribution of the potential scattering. Numerical results of neutron elastic scattering show that 2 orders of Legendre polynomials give DPA rates of $^{56}$Fe within 0.5% overestimation for fission reactors, while 4 orders are required for fusion reactors. For neutron inelastic scatterings, only the first order Legendre polynomial is sufficient to compute DPA rate for both fission and fusion reactors.

## Acknowledgments

The authors acknowledge Dr. Yannick Peneliau for providing the ITER neutron spectrum.